\documentstyle[aps,prb,epsf,rotate]{revtex}

\newcommand\be{\begin{equation}}
\newcommand\ee{\end{equation}}

\begin{document}
\twocolumn[\hsize\textwidth\columnwidth\hsize\csname
@twocolumnfalse\endcsname

\title{Electron/Nuclear spin domain walls in quantum Hall systems} 


\author{Aditi Mitra$^{1,2}$ and S. M. Girvin$^{1}$}
\address{$^1$ Yale University, Physics Department, Sloane Physics Lab, 
New Haven, CT 06520-8120\\
$^2$ Physics Department, Indiana University, Bloomington, Indiana 47405-7105}
\date{\today}
\maketitle

\begin{abstract}

Motivated by recent all optical NMR experiments\cite{Awschalom1,Awschalom2} 
on GaAs quantum wells,
we propose new experiments that would 
involve creating spatially modulated nuclear spin profiles.
Due to the hyperfine coupling 
these would appear as spatially modulated Overhauser fields for the
electrons that could have an amplitude 
large enough to cancel or even reverse
the external Zeeman field at some places.
We discuss 2D electron gas transport in the quantum Hall
regime at filling factor $\nu = 1$, and demonstrate the existence of
collective modes and topological excitations induced in the electron gas
by various nuclear spin patterns. We calculate the 1/T$_1$ relaxation rate 
of the nuclear spins due to coupling with these low lying collective modes
and also discuss how transport and the low
energy modes would be affected by a highly anisotropic g-tensor,
which is special to a GaAs quantum well grown in the 
$\lbrack 110 \rbrack$ direction.

\end{abstract}


\vskip2pc]
 
\section{Introduction}
Quantum Hall systems that are realised in GaAs quantum wells and 
heterostructures show very interesting physics both in terms of 
electron charge and spin degrees of freedom \cite{DasSarma,Girvin}.
Spin orbit coupling
in GaAs reduces the bulk g factor from 2 to -0.4 and together with 
a reduced electron effective mass (m$^*$ $\sim$ 0.07 m$_e$), causes the
Zeeman splitting g$\mu_B$B to be almost 70 times smaller than the cyclotron
energy. Therefore it is possible to be in a temperature regime where
the electrons are confined to the lowest Landau level (LLL), yet low energy 
spin fluctuations are not completely frozen out. Strong coulomb exchange 
interactions together with the dispersionless kinetic energy of the
electrons make these quantum Hall systems ideal ferromagnets, while 
the low energy spin fluctuations are simply the goldstone modes
of the ferromagnet \cite{Girvin}. 

Besides the electron coulomb energy and the Zeeman energy, a third
energy scale, namely the hyperfine interaction of the electrons and
the GaAs nuclei could play an important role in the physics.
As a result of the hyperfine coupling, a net electron spin polarisation 
acts like an effective
magnetic field B$_e$ for the nuclei and the corresponding energy
shift is referred to as the Knight shift.
Similarly, a net nuclear polarisation
shows up as an effective additional magnetic field B$_n$ seen by the
electrons, and the corresponding electron energy shift is referred to 
as the Overhauser shift. In a uniform quantum Hall ferromagnet the
goldstone modes have a gap equal to the Zeeman splitting. This energy
scale while small, is still several orders of magnitude 
larger than the nuclear spin precession frequency 
and therefore the goldstone modes do not affect the
nuclear spin lattice relaxation rates. 


The situation would be different however if 
one could make domain walls in  the electron spins by creating 
regions where the effective electron Zeeman energy changes sign.
We propose this be done by
creating spatial patterns in the degree of nuclear polarization.
Domain walls are characterised by electron spins 
non-colinear to the external applied magnetic field. Due to this, 
the domain walls can support gapless goldstone modes associated
with the zero energy cost for rotating spins around the Zeeman axis.
These zero energy modes can 
effectively couple to the nuclei causing shorter spin lattice 
relaxation times and can also significantly alter charge transport.

Spatial patterns in nuclear polarisation can in principle 
be created by magnetic resonance imaging (MRI) techniques.
Nuclear spins can also be polarised by optical pumping techniques 
\cite{Barrett,Barrett2} and
more recently ultrafast optical spectroscopy \cite{Awschalom1,Awschalom2}
has made it possible to create and detect spatially localised regions of 
polarised nuclear spins, the size of the localised pockets in 
present experiments being 
of the order of 70 microns. This is achieved by focussing the pump
and probe beams by means of a lens. Near field scanning optical
microscopy \cite{Levy} could in principle be used to achieve spatial 
resolutions below the diffraction limit, but has practical difficulties
due to loss of incident light intensity \cite{Kikkawa}.
If this difficulty is
overcome, NSOM would be a very powerful and novel technique to ``write''
various nuclear spin patterns similar to the ones we discuss 
in this paper.

	In this paper we discuss electron transport and
low lying excitations in a quantum Hall system at Landau level
filling factor $\nu=1$ formed in a GaAs
[110] quantum well where the effective Zeeman field seen by the 
electrons, which is
the sum of the external quantising magnetic field and the Overhauser field,
has been spatially modulated.
The spatial modulation may be produced either by near-field scanning 
probe methods or by allowing the pump beam to be a standing wave. 
The latter would create 
a sinusoidal variation in the incident light intensity, and this would 
in turn create a commensurate sinusoidally varying nuclear polarisation. 
From recent experiments \cite{Awschalom1,Awschalom2} we may also assume that 
the nuclear polarisation amplitude is 
large enough to produce oscillations in the sign of the effective Zeeman field
and create lines along which the effective magnetic field seen
by the electrons is zero. 

The hyperfine interaction between the GaAs nuclei and the electron
gas may be written as \cite{Slichter} 
\begin{equation}
H_F = \frac{8\pi}{3}\gamma_e \gamma_n \hbar^2
\sum_{i,j} {\bf S}_i \cdot {\bf I}_j \delta({\bf r}_i - {\bf R}_j)
\end{equation}
Averaging the above expression with respect to the Slater determinant 
state describing a spin polarised 2DEG at $\nu = 1$, it is easy to see
that the Knight shift energy and the Overhauser shift energy are
related as follows when the nuclei carry spin 3/2,
\begin{equation}
\frac{E_{O}}{E_{K}} = \frac{3 p n_{N}}{n_e}
\end{equation}
where $n_e$ and $n_N$ are the 3D electron and nuclear spin densities respectively, 
while p is the extent of nuclear spin polarisation.  
GaAs has a zinc-blend structure with a cubic unit cell side of 5.65 $\AA$,
corresponding to a Ga nuclear density
$n_N = 2.2 \times 10^{28} m^{-3}$. Moreover the size of Knight 
shifts for the Ga nuclei obtained by Barrett {\it et al.} \cite{Barrett2}
is $\sim$ 20kHz. 
Using this and the fact that Knight shifts are enhanced by narrower
wells and higher 2D densities \cite{Sinova}, we estimate the 
Overhauser shift for the samples used by Awschalom's group to be around 
E$_0 \sim 13.2 p $GHz from the Ga nuclei alone. 
Precise calculations \cite{Paget} predict that the nuclear fields
due to polarised As nuclei would be almost twice as large as due to 
Ga, and therefore the maximum
Overhauser shift can be as large as 39.5 GHz,
which corresponds to nuclear magnetic fields (for $g^* \sim 0.053$) of
53 T. Thus by creating nuclear spins polarised in the appropriate 
direction one may have regions where the effective magnetic field seen 
by the electrons is zero or even negative.
      
\section{Domain Walls}
Falko {\it et al.} \cite{Falko1,Falko2} have
described low energy excitations in a quantum Hall system where the 
Zeeman field abruptly changes sign, causing the formation of domain walls
in the electron spin.
They were looking
at the effect of pressure inhomogenities that would cause the g factor
to fluctuate about zero in a sample.
We do a similar analysis to derive the low energy excitations for  
a linearly varying effective Zeeman field which however has been produced
by a controlled spatial manipulation of nuclear spins.  
In addition, we also address the
question of edge transport in a single domain wall system which may 
be a part of an array of domain walls separated roughly by the wavelength of 
the pump beam. 

Fig. 1 is a schematic picture of the domain wall profile.
We will assume that both the  quantising magnetic field that tunes the 2DEG 
to be at $\nu=1$ and the Overhauser field due to the
polarised nuclei point in the 
{$\hat z$} direction. Moreover the spatial variation of the effective Zeeman
field is assumed to be linear and along the {$\hat x$} direction.
The energy density functional describing the 2DEG can therefore be written as 
\cite{Girvin}    
$
H = \frac{\rho_s}{2} \left(\partial_{\mu} m^a \right) \left( \partial_{\mu} m^a \right)
- \frac{e_z}{4 \pi l^2} Q x m^z
$,
where {$\ell$} is the magnetic length, $\rho_s$ is the spin stiffness, 
and $\frac{m_\mu}{2\pi l^2}$ refers to
components of the electron spin density and $e_z$ is the amplitude of the 
effective Zeeman energy which is modulated at wave vector Q.
The orientation of the local electron spin density is described by 
polar angle $\theta(x,y)$ and azimuthal angle
$\phi(x,y)$. In terms of these angles we may write 

\begin{eqnarray}
H &=& \frac{\rho_s}{2} \left [\left(\frac{\partial \theta(x)}{\partial x}
\right )^2 
+ \sin^2\theta(x) \left(\frac{\partial \phi(y)}{\partial y}\right )^2 \right ]
\nonumber\\
&&- \frac{e_z}{4 \pi l^2} Q x \cos\theta(x)
\end{eqnarray}


The above energy functional is minimised by a domain wall solution
where $\phi$ is 
uniform and arbitrary (reflecting the U(1) symmetry in the problem associated
with rotations about the axis of the effective magnetic field), while 
$\theta(x)$ may be chosen to have the 
following variational form \cite{Falko1,comment1}

\begin{equation}
\cos{\theta_0(x)} =  
{\rm sgn}(x)\left [ 1 - 2 {\rm sech}(\beta |x| + \ln(\sqrt2 +1)) \right ]
\end{equation}
The constant term in the argument of sech ensures the boundary condition 
that for $\beta x = -(+)\infty$, $\theta = \pi$(0) and at $x=0$, 
$\theta = \frac{\pi}{2}$.
We determine $\beta$ by requiring that 
$\frac{\partial}{\partial \beta} \int dx H(x,\beta) = 0$ and find 
$
\beta = \left [ \frac{2 \sqrt2}{\sqrt2 -1} \ln{(\frac{2\sqrt2}{\sqrt2 + 1})} 
\frac{e_z}{4 \pi l^2 \rho_s} Q \right]^\frac{1}{3}
$.
  
We will now proceed to derive an effective action describing the spin
wave modes that arise when $\theta$ and $\phi$ fluctuate about $\theta_0$ 
and $\phi_0 = $ constant respectively.
The leading order term in the 
effective energy functional for the spin waves has two parts. 
One part is obtained by simply integrating out the the domain wall profile in
the {$\hat x$} direction in the second term in Eq 3. 
This gives the following term in the action 
\begin{equation}
U \approx \frac{\rho_s}{2}\frac{1.72}{\beta} \int dy \left 
(\frac{\partial \phi}{\partial y} \right )^2 
\end{equation}
 
The second term in the spin-wave energy functional is a measure of the Zeeman energy cost
for deviations from $\theta_0$  
and is obtained variationally by evaluating the change in Zeeman energy for a domain wall
profile whose center is shifted from $x=0$ to $x=X_0$. This gives rise to a 
net magnetisation 
density along the transverse {$\hat y$} direction $m_z^{1D} = \frac{X_0}{\pi l^2}$ and the 
cost in Zeeman energy turns out to be 
\begin{equation}
E_z = \int dy \, \frac{\pi e_z Q l^2}{4} \left (m_z^{1D}(y) \right )^2
\end{equation}

The action may be derived by starting with the Berry's phase term for a 
quantum Hall ferromagnet
\cite{Girvin}  $ S n\int dx \int dy A^{\mu} \dot{m_{\mu}} $, 
where $\vec{A} = \vec{\nabla}_m \times \vec{m} $.  For the domain wall region, 
the spin is almost 
completely in the x-y plane and this would mean $ \vec{A} \approx -m_z \hat{y}$.  
Using this and 
integrating out the {$\hat x$} direction, the Berry phase term for the 1D action is
$\frac{1}{2} \int dy \dot{\phi} m^{1D}_z$.  

Combining all these terms, the total action describing the domain wall is 
\begin{equation} 
S = \int dy \int d\tau \, \frac{i}{2} \frac{\partial \phi}{\partial \tau} m^{1D}_z 
- \frac{\Gamma}{8} (m^{1D}_z)^2 - \frac{\rho_s^{1D}}{2} \left 
( \frac{\partial \phi}{\partial y} 
\right )^2\end{equation}
where $\Gamma = 2\pi l^2 Q e_z$ and $\rho_s^{1D} = \frac{1.7}{\beta}\rho_s$.
Next, one can integrate out the massive $m_z^{1D}$ fluctuations 
from the above action and arrive at 
$S = \int dy \int d\tau \, \frac{\rho_s^{1D}}{2} \left 
( \frac{\partial \phi}{\partial y} \right )^2 
+ \frac{1}{2\Gamma} \left ( \frac{\partial \phi}{\partial \tau} \right )^2  $.

The quantised Hall conductance implies a direct relation between the charge and
spin fluctuations \cite{Girvin}
$\delta\rho(x,y) = \frac{1}{8\pi}\epsilon^{\alpha \beta} \vec{m} \cdot 
(\partial_{\alpha} \vec{m} \times \partial_{\beta} \vec{m})$.
For the domain wall profile which minimises the energy functional, integrating over the 
{$\hat x$} direction gives the following relation for 1D
charge deviations along the {$\hat y$} direction
$\rho(y) = \frac{1}{2\pi}\frac{\partial \phi}{\partial y}$.
Thus the above action for spin waves along the domain wall can be mapped onto a Luttinger 
liquid with interaction parameter $g =\frac{1}{4 \pi}\sqrt{\frac{\Gamma}{\rho_s^{1D}}} $ and 
collective mode velocity $ c = \sqrt{\Gamma \rho_s^{1D}}$.     

Assuming that the nuclear spin profile is established optically by either using 
a standing light wave of wavelength $\lambda$ or by a diffraction limited
focussed beam, the scale of Q will be set by $\lambda = \frac{2\pi}{Q}\sim 100 \ell$.
The above estimate for Q,
along with the fact that $e_z \sim 3$K and using \cite{Moon}
$\rho_s = \frac{1}{16 \sqrt{2\pi}}\frac{e^2}{\epsilon \ell}$ gives a domain wall width,
$\frac{1}{\beta} \approx 6\ell$. 
For this the collective mode velocity along the 
domain wall is estimated to be $\sim$ 10$^3$ m/s.

As mentioned before, the spin waves along the domain wall are gapless and therefore
can couple to the GaAs nuclei causing a 
finite spin-lattice relaxation time T$_1$ given by 
\begin{equation}
\frac{1}{T_1} = \lim_{\omega \rightarrow 0}
\frac{2\pi}{\hbar}(\frac{A}{2})^2 |<\frac{3}{2}|I_+|\frac{1}{2}>|^2
\frac{2}{1-e^{-\beta \omega}}{\rm Im}\chi_{+-}(\omega)
\end{equation}
where Im$\chi_{+-}(\omega)$ is the dissipative part of the spin 
susceptibility. Note that A is related to the Knight Shift as
$E_K = A <S_z>$.
In our formalism, the spin raising operator is defined as
$S_{+}(x,y,\tau) = \frac{1}{2}\sin\theta_0(x) e^{i\phi(y,\tau)}$ where y is a coordinate
along the domain wall, while $x$ is the coordinate in the transverse direction.
Im$\chi_{+-}$ is
obtained by evaluating the
following correlation function in imaginary time, $ C(\tau) =  
<T_{\tau}S_+(x,y=0,\tau) S_-(x,y=0,0)> $ , 
followed by simultaneously doing the fourier 
transform and the analytic continuation using the identity \cite{Balents}
Im$\chi_{+-}(y=0,\omega) = 
\sinh{\frac{\beta \omega}{2}} \int dt e^{-i\omega t} C(\beta/2 -it)$.
Due to the gaussian action of the Luttinger liquid, the above 
correlation function can be analytically evaluated and we obtain
\begin{equation}
\frac{1}{T_1} =\frac{\sin^2\theta_0(x)}{T^{1-2g}}
\frac{2\pi}{\hbar}(\frac{A}{4})^2
|<\frac{3}{2}|I_+|\frac{1}{2}>|^2 \frac{(2\pi)^{2g-1}}
{(\Lambda v_F)^{2g}}\frac{\Gamma^2(g)}{\Gamma(2g)}
\end{equation} 
The above expression yields the 
the Korringa law T$_1$ T $=$  constant in the non-interacting limit $g=1$. 
Note that $\Lambda v_F$ is a short time cutoff that arises in evaluating the
correlation functions. While we can only give a heuristic estimate for
what this cut-off ought to be, we find from the variational treatment of
the domain wall above that $g \sim 0.02$. For such tiny values of g
the size of T$_1$ is not very sensitive to our choice of cut-off. The
reason why g is much smaller 
than is typical for 1D interacting electron
gases is that for our system g scales as $g \sim (Ql)^{\frac{1}{3}}$, where
Q is the wave-vector of the pump beam.  
If we assume the momentum cutoff to be set by inverse of the domain wall
width ($\Lambda \ell \sim \frac{1}{6}$)
and if we take $v_F \sim 10^3$m/s, which yields an energy cutoff 
($\hbar \Lambda v_F$) of 1K,  we find T$_1$ $\sim$ 0.1s at the center of the 
domain wall and at a temperature of 1K (for samples similar to those used
in Ref. 1 and 2).
This time scale indicates that the domain
wall will stay intact long enough to carry out electron transport measurements.
This time scale is however shorter by at least 4 orders of magnitude
from typical nuclear relaxation times observed in quantum Hall samples in uniform Zeeman
fields close to $\nu = 1$, and 2 orders of magnitude smaller than nuclear 
relaxation times observed in  
$\nu = 0.88$ quantum Hall samples that contain skyrmions \cite{Barrett2}. 

It is interesting to notice that for times longer than T$_1$, the nuclei at $x=0$ would
depolarise significantly. 
But since the nuclear spins far from the domain wall center are still polarised
parallel/anti-parallel to the external Zeeman field, the domain wall in electron spins
and the gapless excitations characterised by them will remain intact. However the
details of the domain wall profile (length, spin wave velocity etc.) would get 
modified and the center of the domain wall will also shift along the direction
of the nuclear field gradient.

We now address the question of transport in the above geometry. 
Figure 2 shows what we 
have in mind. We imagine feeding current into the sample through the edges 
that are
adiabatically connected to the reservoirs at the two ends, while towards the 
center of the 
sample the spin degree of freedom associated with the edges rotates to 
follow the
domain wall profile calculated above. We would like to know the 
transmission probability
across this domain wall. Perfect transmission would mean that voltage 
probes V1 and V2 are at the same potential (i.e., $\rho_{xx} = 0$) and the $\nu=1$
quantum Hall state is restored. Perfect reflection on the other hand would correspond
to destruction of the quantum Hall state \cite{MitraGirvin}.

In order to analyse this we note that the domain wall is equivalent to
two interacting chiral Luttinger liquids of {\it opposite} spins. This is
made clear by doing the following change of variables in the domain wall 
action (Eq. 8),
$\phi = \phi_{\uparrow} + \phi_{\downarrow}$ and 
$m_z = \frac{1}{2\pi}(\frac{\partial \phi_{\uparrow}}{\partial x} - 
\frac{\partial \phi_{\downarrow}}{\partial x})$. This change of variables preserves the 
canonical commutation relations between $\phi$ and $m_z$ , provided $\phi_{\uparrow}$
and  $\phi_{\downarrow}$ obey the Kac-Moody algebra. In terms of these new variables 
the action may be written as S = S$_L$ + S$_R$ + S$_{\rm int}$
where 
\begin{equation}
S_{L/R} = 
\int dx \int dt \frac{1}{4\pi}(\frac{\partial \phi_{\uparrow}}{\partial x})
({(\pm)}\frac{\partial \phi_{\uparrow}}{\partial t} - v_0 \frac{\partial \phi_{\uparrow}}{\partial x})
\end{equation}
\begin{equation}
S_{\rm int} =  -\int dx \int dt \, \lambda (\frac{\partial \phi_{\uparrow}}{\partial x})
(\frac{\partial \phi_{\downarrow}}{\partial x}).
\end{equation}
and $\lambda = \rho_s^{1D} - \frac{\Gamma}{16\pi^2}$ and $v_0 = 2\pi\rho_s^{1D} + 
\frac{\Gamma}{8\pi}$.
S$_L$ and S$_R$ are actions for left-moving and right-moving non-interacting 
chiral Luttinger liquids while the third term shows that these chiral
modes interact with each other through the parameter $\lambda$. This also exlains
why the spin raising operator S$_+ \propto e^{i\phi}$. This is because the spin raising
operator is equivalent to $\Psi_L^{\dagger}\Psi_R \propto e^{2 i k_F y} 
e^{i(\phi_{\uparrow} + \phi_{\downarrow})}$. Note that for our problem $k_F$ is zero
since microscopically the domain wall profile is obtained by taking a linear 
combination of up-spin and down-spin single electron states at the same 
momentum (or guiding center index). For the usual Luttinger liquid obtained 
from bosonizing spinful fermions, the spin raising operator
has a more complicated form 
since unlike our case, a spin flip can occur along more than
one channel (such as spin flips between unidirectional {\it and} 
counterpropagating modes).

We can now describe transport via the edge modes in Fig. 2 as the same 
as transport 
along a 1D chain that has 3 parts. The 1st and 3rd parts consist of Luttinger liquids
with interaction parameter $g=1$. This represents the chiral non-interacting edge-modes that
feed in/out of the reservoirs. The middle part of the chain however is an interacting
Luttinger liquid and represents the domain wall. 
Transport in such coupled Luttinger liquids
have been studied by various authors \cite{Safi,Oreg}. The central result 
of Safi {\it et al} \cite{Safi}  is that as long as there is no disorder, 
the central wire acts as a Fabry 
Perot interferometer and there is always perfect transmission along the wire in the dc limit.
In the present problem perfect transmission occurs because of 
conservation of the $z$ component of the electron spin and due to the fact
that the two chiral modes carry opposite spins, which makes backscattering impossible.
Perfect transmission along the 1D chain
implies that a wave incident along one of the edges 
in Fig. 2 travels along the domain wall and gets perfectly reflected back into the 
same reservoir. In our language therefore this corresponds to perfect reflection
at the domain wall and this would give rise to a finite voltage drop between probes 
V1 and V2 and hence a destruction of the $\nu=1$ quantum Hall state.

One could get finite transmission  across the domain wall if the spin waves 
along the domain wall are gapped. When this happens a low energy
mode incident from one of the reservoirs would not have any propagating 
state to scatter to along the domain wall. Thus it would travel along the domain
wall as an evanescent wave, and for a long enough domain wall,
may completely decay before reaching the other end. This situation would correspond
to complete transmission across the domain wall.
The physics of this has been analysed in detail \cite{MitraGirvin} in a different 
context involving two $\nu=1$ 2DEGs separated by a narrow but high barrier.
In the next section we discuss how such a gapping can arise.

\section{Anisotropic g-tensor}
In our present set-up the spin waves can be gapped by introducing spin-orbit
interactions which destroys the U(1) symmetry for rotations about the Zeeman axis.
The effect of spin-orbit interactions in the regime of vanishing
Zeeman energy has been studied in detail by Falko {\it et al.} \cite{Falko1,Falko2}.
They explicitly show \cite{Falko1} that a Rashba spin-orbit interaction gives rise to
a small additional 
term in the spin wave action proportional to $\cos\phi$, so that the total action looks 
like the integrable Sine-Gordon model.  

We find that for a 2DEG formed in a GaAs [110] heterostructure   
the spin waves can be gapped even in the absence of explicit spin-orbit
coupling terms. This is due to the anisotropic g tensor (implicitly due to 
spin-orbit effects) and may be understood as follows.
The crystal symmetry in GaAs is such that the principal axes of the 
g-tensor coincide with the 
[110],[-110] and [001] directions. For orientations of the external B 
field that do not coincide with the principal axes, the electrons spins
would want to align along an axis $\Omega^\mu = g^{\mu \nu}B_{\nu}$, 
noncollinear to {\bf B}.
Unlike the electrons, the polarised nuclear spins would continue to
precess about {\bf B} and therefore 
the time-averaged nuclear spin points along {\bf B} and may be written as 
$<{\bf I}> = I({\bf r}) {\bf \hat B}$.
As the nuclear polarisation $I(x)$ varies spatially, the total effective
magnetic field now rotates in the {\bf $\Omega$} - {\bf B} plane.
This is a more complicated situation than before where the net B field was
always along the same axis. 
If we denote $\phi$ to be the angle that the electron spin makes 
with the {$\bf \Omega-B$} plane, and $\theta$ the angle it makes with 
{\bf $\Omega$}, then the Zeeman energy density has the following form
\begin{eqnarray}
E_z &=& s_{\parallel}\left( \Omega + 
E_O p(x) {\bf \hat \Omega} \cdot {\bf \hat B}\right) \nonumber\\
&& + s_{\perp}E_O p(x) |{\bf \hat \Omega} \times {\bf \hat B}| \cos\phi
\end{eqnarray}
s$_{\parallel/\perp}$ is the component of the electron spin density 
along/perpendicular
to {\bf $\Omega$} and $E_O$ is the amplitude of the Overhauser shift
and $p(x)$ is the spatially varying net nuclear spin polarisation 
along {\bf B}.

For small deviations of the applied magnetic field from the crystal 
symmetry axis, one can still assume that the domain wall profile 
is given by Eq. 4.  
By looking at fluctuations about the static domain wall solution one
now obtains the integrable Sine-Gordon model, characterised by a Luttinger
liquid part and  an additonal term 
proportional to $\cos\phi$. The coefficient of
the $\cos\phi$ term reflects the fact that 
the spin waves are gapped iff there is a net nuclear
polarisation in addition to an anisotropic g-tensor. The spin wave gap 
keeping only the leading order term in $|\bf \hat \Omega \times \bf \hat B|$ is 
given by $\Delta_s = \sqrt{\Gamma \frac{\Omega}{4\pi l^2}
(\int dx \sin\theta_0(x))
|{\bf \hat \Omega} \times {\bf \hat B}|} \approx 0.66
\sqrt{|\bf \hat \Omega \times \bf \hat B|}$ K for a 10 T field slightly
misaligned with the [110] axis. Transport measurements are sensitive
to the charge gap ($\Delta_c$) which is twice the energy required to 
create a soliton in the field $\phi$, and the classical expression for this
to leading order in the tilt angle is given by
$\frac{\Delta_c}{2} = \frac{2}{\pi}\frac{\Delta_s}{g} \approx 26.3 
\sqrt{|\bf \hat \Omega \times \bf \hat B|}$ K. This 
estimate for the charge gap is in principle reduced by quantum
fluctuations which may be calculated exactly \cite{Kollar}, but 
in the limit $g \rightarrow 0$, the classical estimate is more
and more exact \cite{MitraGirvin}. Therefore for the purposes here
($g \sim 0.02$) we use 
the classical expression and
find that at temperatures of 1K, for the
domain wall excitations to appear gapless, the external B field has to
be aligned along [110] as precisely as $0.08 \deg$. The evanescent 
wave decay length ($\frac{1}{K} = \frac{\hbar c}{\Delta}$)
at this angle is only 77\AA \, which also sets the upper limit
on the domain wall length for which one would observe perfect 
reflection at the domain wall.
Due to these two reasons, namely the precision
with which the magnetic field has to be aligned along the the
[110] crsytal symmetry axis and the difficulty in creating domain walls
of length $<\sim \ell$, experimentally it seems one would always
measure a finite charge gap. 

\section{Skyrmions}
	It is also interesting to study the nature of the collective modes in a 
geometry where a tiny circular patch of region has spin-reversed nuclei, yielding 
an effective B field which 
is along the negative $\hat z$ direction, while everywhere outside the patch
the B field is
along the positive $\hat z$ direction. One would now expect the domain wall of spins
to be circular (see Fig. 3).
Let us suppose the radius of the patch is R. Then the excitations
along the domain wall will be of two kinds. The first will be neutral 
excitations that do not carry any topological charge,
and have an energy given by $\omega_n = c k_n = 
\sqrt{\Gamma \rho_s^{1D}}\left(\frac{n}{R}\right)$, where n is 
any integer. The second kind of excitation along the domain wall are charge carrying
excitations where the field $\phi$ winds by 2$\pi$m along the circumference and
therefore carries a
net charge of $m e$. The energy of these modes which is a sum of
the exchange energy and the electrostatic hartree energy of a ring of charge
of radius R, in the limit of $r_0 << R$, is given by
$\omega_{m} = \frac{m^2}{R}\left(\pi \rho_s^{1D} - \frac{e^2}{2 \pi \epsilon}
\ln\frac{r_0}{8 R} \right)$,
where m now labels the topological charge, 
and $r_0 \sim \ell$ is an ultraviolet cutoff associated with the finite thickness of
the ring of charge.

	Skyrmions are charged topological spin excitations that arise in quantum Hall 
ferromagnets
and their energy and size is determined by a competition between coulomb interactions
and the Zeeman energy.  
Detailed Hartree-Fock calculations have been done
\cite{Fertig} that estimate the size of
a skyrmion in GaAs to be a few magnetic lengths so that the skyrmions
contain about 3-4 overturned spins. NMR Knight shift experiments also
\cite{Barrett2} support this estimate. However skyrmions can be much larger
at high pressure where the g factor is reduced \cite{Maude} and at $\nu=1$
can have arbitrarily many flipped spins in the limit  $g^* \rightarrow 0$. Here 
we point out that instead of pressure tuning, one could vary the degree of nuclear
polarization to obtain small effective Zeeman fields and hence large scale size
skyrmions. In addition, by spatially modulating the degree of nuclear polarisation one
could produce effective potential wells that could trap the skyrmions and/or
modify their transport. 
For example, if the regions inside and outside the circular patch
are characterised by an effective Zeeman field of the same strength, but
opposite orientations, then skyrmions with K spin flips
formed outside the patch (say) would be attracted to the patch which would
appear as a potential well of depth $2 g^* \mu_B B K \sim 2.2$ Kelvin for 
a 10 T field and $g^* \sim 0.04$. 
This situation is not stable however and for time scales longer than 
T$_1$ the skyrmion would increase in size by acquiring additional spin flips.
Eventually the
skyrmion would turn into the topological excitation described above (Fig. 3)
whose charge lies on a circular ring domain wall and the electron spins in 
the interior of the ring would be all reversed. 

\section{Bilayer Domain Walls}
	Analogous to the derivation done above for the domain wall 
collective modes in a $\nu=1$ quantum Hall sample, we can do a similar
analysis for a quantum Hall double-layer \cite{Eisenstein}, 
where the spins are replaced by pseudospins \cite{Girvin,Moon}. The pseudospin  
labels which layer the electron is in
and the spatially varying effective Zeeman field is replaced by a spatially
varying external bias potential which unbalances the charge density in the 
two layers. With appropriate split gates one could arrange the external bias 
potential to be large and 
positive for $x<0$ say, and large and negative for $x>0$, so that the 
pseudospin
is ``up'' on the left and ``down'' on the right, {\it ie}, the electrons like
to sit completely in the upper/lower layer on the left/right. 

	In the intermediate region around $x=0$ we would however expect
a pseudospin domain wall where the pseudospin gradually tilts from +1/2 
to -1/2 and in this region the electrons may be regarded to
be in a coherent superposition of both layers \cite{Girvin,Moon,Spielman}.
We do a  similar analysis as before on the full energy functional
for the double layer given by \cite{Moon}
\begin{eqnarray}
E[{\bf m}] &=& \int d^2r \beta (m_z - V_B(x))^2 + C[{\bf m}] + 
\frac{\rho_A}{2} (\nabla m_z)^2 \nonumber\\
&& + \frac{\rho_E}{2} [(\nabla m_x)^2 + (\nabla m_y)^2]
\end{eqnarray}
$V_B(x)$ is the external bias potential which we assume varies linearly and passes
though zero at $x=0$.
The effective action for the
domain wall we derive also looks like that of a Luttinger liquid, where the 
scalar field $\phi$ is now related to the X-Y orientation of the pseudospin
along the domain wall.
For $\beta = 0.005 \frac{e^2}{\epsilon l^3} $ and $\rho_E = 
0.012 \frac{e^2}{\epsilon l} $, the 
velocity of the mode turns out to be $v \sim 0.17 \frac{e^2}{\epsilon} = 
3 \times 10^4$ m/s, slightly larger than the bulk collective mode velocity
in a balanced double layer system \cite{AdyStern}.

\section{Conclusion}
	The method of all optical NMR allows the nuclear
polarisation to be spatially modulated.
Moreover tiny and anisotropic
g-tensors can give rise to large Overhauser fields that can cancel the external
Zeeman field seen by the electrons. In this paper we have
discussed electron transport
and the nature of low-lying excitations for a variety of spatial patterns of 
nuclear spin polarisations that we suggest can now be achieved
experimentally.

\section{Acknowledgements}
We would like to thank D. D. Awschalom and J. P. Eisenstein
for helpful discussions and useful suggestions.
This work was supported by NSF DMR-0087133.     

\begin{figure}
\epsfxsize=2.6in \centerline{\epsffile{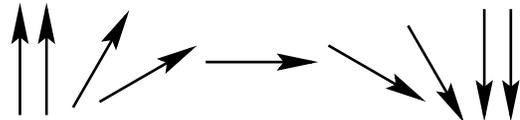}}
\caption{Electron spin domain wall profile in the vicinity of vanishing
effective Zeeman field.} 
\end{figure}

\begin{figure}
\epsfxsize=2.6in \centerline{\epsffile{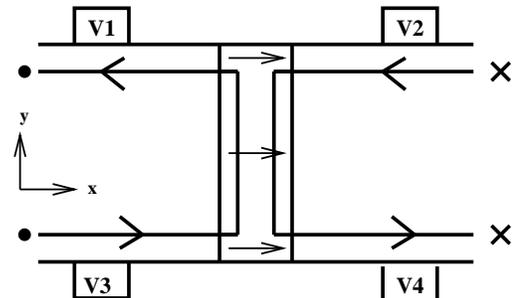}}
\caption{Top view of sample. Central portion is 
the domain wall characterised by electron spins with an in-plane
component. In the absence of spin-orbit interactions, S$_z$ is
conserved. Hence edge modes are completely reflected at the 
domain wall.} 
\end{figure}

\begin{figure}
\epsfxsize=2.6in \centerline{\epsffile{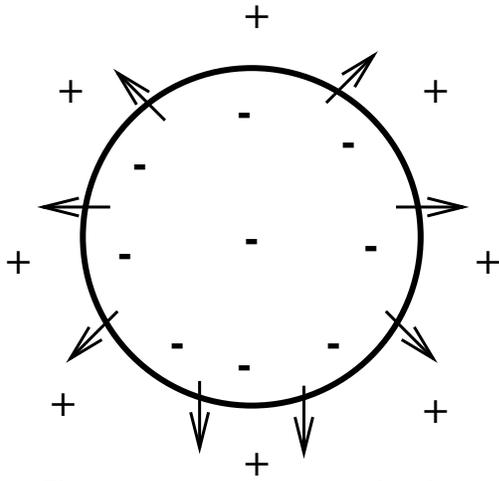}}
\caption{Electron spin texture associated with creating circular
pockets of nuclear spins polarised anti-parallel to the external
applied field. Here the charge 
associated with the circular domain wall is exactly $1e$.} 
\end{figure}


\begin{references}

\bibitem{Awschalom1} G. Salis, D. D. Awschalom, Y. Ohno and H. Ohno, 
cond-mat/0104564.

\bibitem{Awschalom2} G. Salis, D. T. Fuchs, J. M. Kikkawa, and D. D. Awschalom,
{\sl Phys. Rev. Lett}, {\bf 86}, 2677 (2001). 

\bibitem{DasSarma} S.M. Girvin, and A.H. MacDonald in {\sl
Perspectives in Quantum Hall Effect}, Ed. S. Das Sarma and A.
Pinczuk, John Wiley and Sons, Inc.(1997).   

\bibitem{Girvin} S. M. Girvin, `The Quantum Hall Effect: Novel Excitations and
Broken Symmetries,' 120 pp. Les Houches Lecture Notes, in:
{\sl Topological Aspects of Low Dimensional Systems,}
ed. by Alain Comtet, Thierry Jolicoeur, Stephane Ouvry and Francois
David, (Springer-Verlag, Berlin and Les Editions de Physique, Les Ulis,
2000), (eprint: cond-mat/9907002).                 

\bibitem{Barrett} S. E. Barrett, R. Tycko, L. N. Pfieffer, and K. W. West,
{\sl Phys. Rev. Lett}, {\bf 72}, 1368 (1994).

\bibitem{Barrett2} S. E. Barrett,G. Dabbagh, L. N. Pfieffer, K. W. West,
and R. Tycko, {\sl Phys. Rev. Lett}, {\bf 74}, 5112 (1995).

\bibitem{Levy} J. Levy, V. Nikitin, J. M. Kikkawa, A. Cohen, N. Samarth,
R. Garcia, and D. D. Awschalom, {\sl Phys. Rev. Lett.}, {\bf 76},
1948 (1996).

\bibitem{Kikkawa} J. M. Kikkawa, private communication.

\bibitem{Slichter} C. P. Slichter, {\sl Principles of Magnetic Resonance} 
(Harper and Row, New York, 1963).

\bibitem{Sinova} J. Sinova, S. M. Girvin, T. Jungwirth, and K. Moon, 
{\sl Phys. Rev. B.} {\bf 61}, 2749 (2000).

\bibitem{Paget} D. Paget, G. Lampel, B. Sapoval, and V. I. Safarov, 
{\sl Phys. Rev. B}, {\bf 15}, 5780 (1977).

\bibitem{Falko1} V. I. Falko and S. V. Iordanskii, {\sl Phys. Rev. Lett.} 
{\bf 82}, 402 (1999).

\bibitem{Falko2} V. I. Falko and S. V. Iordanskii, {\sl Phys. Rev. Lett.} 
{\bf 84}, 127 (2000).

\bibitem{comment1} The chosen variational form was shown by Falko {\it et al}
to satisfy the saddle point equation for $\theta$ when the Zeeman field 
behaves like a step function at $x=0$. Other variational forms should yield
quantitatively similar estimates for the domain wall width.

\bibitem{Moon} K. Moon, H. Mori, Kun Yang, S. M. Girvin, and A. H. MacDonald,
{\sl Phys. Rev. B}, {\bf 51}, 5138 (1995).

\bibitem{Balents} L. Balents, 1999 Moriond Les Arcs 
Conference Proceedings, (eprint: cond-mat/9906032).

\bibitem{MitraGirvin} Aditi Mitra and S. M. Girvin, {\sl Phys. Rev. B}, {\bf 64}, 041309(R)
(2001).

\bibitem{Safi} I. Safi and H. J. Schulz, {\sl Phys. Rev. B}, {\bf 52}, R17040 (1995).

\bibitem{Oreg} Y. Oreg and A. M. Finkelstein, {\sl Phys. Rev. B}, {\bf 54}, 14265 (1996).

\bibitem{Kollar} M. Kollar and S. Sachdev, cond-mat/0106001.

\bibitem{Fertig} H. A. Fertig, L. Brey, R. C\^{o}t\'{e} and A. H. MacDonald, 
{\sl Phys. Rev. B}, {\bf 50}, 11018 (1994).

\bibitem{Maude} D. K. Maude, M. Potemski, J. C. Portal, M. Henini, L. Eaves,G. Hill, 
M. A. Pate, {\sl Phys. Rev. Lett}, {\bf 77}, 4604 (1996); D. R. Leadley,
R. J. Nicholas, D. K. Maude, A. N. Utjuzh, J. C. Portal, J. J. Harris, 
C. T. Foxon, {\sl Phys. Rev. Lett}, {\bf 79}, 4246 (1997).

\bibitem{Eisenstein} We thank J. P. Eisenstein for suggesting this.

\bibitem{Spielman} L. B. Spielman {\it et al}, {\sl Phys. Rev. Lett},
{\bf 84}, 5808 (2000).

\bibitem{AdyStern} A. Stern, S. M. Girvin, A. H. MacDonald, and N. Ma, 
{\sl Phys. Rev. Lett.}, {\bf 86}, 1829 (2001).

\end{references}
\end{document}